# Symmetrical Graph Neural Network for Quantum Chemistry, with Dual R/K Space

Shuqian Ye,[1,2] Jiechun Liang,[1,2] Rulin Liu,[1,2] Xi Zhu[1,2*]

[1]Shenzhen Institute of Artificial Intelligence and Robotics for Society (AIRS), Shenzhen 518172, China.

[2]The Chinese University of Hong Kong, Shenzhen, Shenzhen 518172, China.

[*]zhuxi@cuhk.edu.cn

## Abstract

Most of current neural network models in quantum chemistry (QC) exclude the molecular symmetry, separate the well-correlated real space (R space), and momenta space (K space) into two individuals, which lack the essential physics in molecular chemistry. In this work, by endorsing the molecular symmetry and elementals of group theory, we propose a comprehensive method to apply symmetry in the graph neural network (SY-GNN), which extends the property-predicting coverage to all the orbital symmetry for both ground and excited states. SY-GNN shows excellent performance in predicting both the absolute and relative of R and K spaces quantities. Besides the numerical properties, SY-GNN also can predict the orbitals distributions in real space, providing the active regions of chemical reactions. We believe the symmetry endorsed deep learning scheme covers the significant physics inside and is essential for the application of neural networks in QC and many other research fields in the future.

## Introduction

The design of molecules with on-demand properties is of significant importance to developing functional materials and drugs, and one of the leading methodologies is density functional theory (DFT)[1]; however, DFT is extremely time-consuming for the large molecules. To accelerate this process, recently, various machine learning (ML) methods have been applied to QC calculations to approximate molecular energies and other properties.[2-14] The main advantage of the ML techniques is that, once the model is well-trained, the predictions can be rapid. Meanwhile, deep neural networks (DNNs) have been applied to achieve fast QC calculations. Compared to the kernel methods in ML, the training time of the DNNs scales linearly with the number of data samples. While some approaches make use of painstakingly designed descriptors[15], message passing algorithms[3], a type of graph neural network, has been applied to predict properties from molecular structures. Very recently, some neural networks that learn a representation directly from atom types and positions are widely applied. Deep Tensor Neural Networks (DTNNs)[12] is one that enables spatially and chemically resolved insights into the energy of molecules. SchNet[5], as a variant of DTNNs, uses continuous filter convolutional layers to learn representations for molecules and materials to achieve high precision property prediction in a wide range of quantum-mechanical properties.

Though the success of ML and DNNs has led to wide-spread applications in other scientific fields, the interpretability, and dataset-independent transferability are essential prerequisites to guarantee the value of the ML/DNNs application outcome, especially in QC. However, the efficiency of some current DNNs are dataset dependent and are not working for predicting large molecules.[16] Both the prediction error, memory cost, and training time will dramatically increase while the size of the molecule increasing, indicating the pure numerical ML/DNNs architecture is well-challenged for the large scale demand from QC. Moreover, till now most of the ML/DNNs architectures are introduced from the areas like computer-science or data-science, in which research areas there is little physics-constrained data correlation, while in QC research, all of the data correlates with the K and R spaces, and a severe defect is that the K and R spaces quantities are treated as separated ones in the previous ML/DNNs' application in QC, which maps the scientific physical chemistry data into pure engineering numerical ones, the solid-state physics or molecular orbital theory actually disappears. The physics-inspired ML models can provide more accuracy and a secondary understanding from the raw data[17]. The critical loss for the K/R correlation is the ignorance of the symmetry, which directly constrains the K/R quantities.

In traditional QC DFT calculation, the symmetry is widely applied to reduce the computation cost of the DFT to that for an N/4 point fast-Fourier-transformation (FFT) with pre-processing and post-processing.[18] The symmetry operation provides an efficient solution for the structure refine and property prediction for some large molecules in the DFT scheme. Here in order to solve the dataset dependent issue in the DNNs scheme, we extend the symmetry operation to the DNNs scheme and present a novel deep learning model SY-GNN, which includes the symmetric atomic interactions and is capable for covering both the R space and K space features. The introduction of symmetry operation in SY-GNN not only just improved the numerical quantity, more importantly, since the symmetry itself dedicates the atomic interaction, but it also endorses the solid-state physics, which makes the SY-GNN acts more robust, especially in the energy degeneracy and excited states selection than other now available DNNs.

**Method**

Fig. 1A shows the mainframe and overall architecture of SY-GNN. First, we define a molecule $\mathcal{M}$ as a set of $n$ atoms, i.e. $\mathcal{M} = \{(Z_1, \boldsymbol{r}_1), (Z_2, \boldsymbol{r}_2), \cdots, (Z_n, \boldsymbol{r}_n)\} = \{(Z_i, \boldsymbol{r}_i)\}_{i=1}^n$, where $Z_i$ is the atomic number, i.e. atom type of the $i$th atom, and $\mathbf{r}_i$ is the position of the atom in the 3-dimensional Cartesian coordinate of the $i$th atom. Though the layers of SY-GNN, the atoms are represented by a tuple of features $X^l = (\boldsymbol{x}_1^l, \cdots, \boldsymbol{x}_n^l)$, where $\boldsymbol{x}_i^l \in \mathbb{R}^F$ with the $l$th layer, the number of atoms $n$, the number of feature map $F$. The feature of site $i$ is initialized using an embedding layer dependent on the atom type $Z_i$

$$x_i^0 = \boldsymbol{a}_{Z_i} \quad (1)$$

These embeddings $\boldsymbol{a}_Z \in \mathbb{R}^d$ are vectors to represent information of atoms disregarding the environment they are in, where d is the dimensionality of vectors. Furthermore, they are initialized randomly and optimized via backpropagation during the training process.

The graph theory layer (GT) is a layer use the related rotation matrix of different symmetry group to find out the primitive unit and the equivalent atoms in a symmetrical molecule. Section S2 shows the details of the GT layer.

Since the molecules are symmetric, the same type of atoms in the symmetric position should have identical features, and we can share tensor values among them but not do the redundant calculation. At the beginning of interaction modules, we pick-up only atoms in a minimal part $\mathcal{S} \subset \mathcal{M}$ of a symmetrical molecule. For each molecule $\mathcal{M}$ with symmetry, there is the symmetrical operation $\Gamma()$ that can be used to perform on the minimal part $\mathcal{S}$, and $\mathcal{M} = \Gamma(\mathcal{S}) \cup \Gamma(\Gamma(\mathcal{S})) \cup \cdots$.

In the interaction modules, we share their features with other symmetric atoms after all calculations in the interaction module. It means that all the atoms belonging to the same symmetry operation share the same features in the output.

$$x_{i \in \mathcal{S}}^{(l+1)} = x_{i \in \Gamma(\mathcal{S})}^{(l+1)} \qquad (2)$$

For a molecule with $D_{6h}$ symmetry like benzene in Fig. 1B, during each turn of interaction calculation, we only need to calculate the interaction between one carbon or hydrogen atom and the other atoms. After updating the features for these two atoms, keep the features of other symmetric atoms and the new feature of the carbon and hydrogen atoms in synchronization. Theoretically, we can reduce the amount of calculation by five-sixths.

In the previous models for predicting quantum chemical properties like DTNNs and SchNet, residual connections[19] are applied to connect each layer. SY-GNN applies the dense-connections scheme[20], and symmetries in molecules are applied to enhance the learning ability of the neural network. Since densely connectivity is used in our architecture, the number of output features of the interaction module is $F$, while the number of input features is $mF$, which is related to the index of interaction module $m$. We concatenate all output from all previous modules to get the input for this module.

$$x^{(m)} = \bigoplus_{i=1}^{m-1} x^{(i)} \qquad (3)$$

which "$\bigoplus$" represents the concatenation of the feature-maps produced in modules *1, 2, …, m-1*.

The first layer in the interaction module is an atom-wise dense layer that used to make dimensionality reduction of the feature map, which means $\sigma: \mathbb{R}^{mF} \to \mathbb{R}^F$. Atom-wise dense layers are fully-connected layers applied to the features $x_i^{(l-1)}$ of each atom *i* in layer *l-1* to learn atom interactions and return the features $x_i^{(l)}$ as the input of the next layer.

$$\sigma: x_i^{(l)} = W^{(l)} x_i^{(l-1)} + b^{(l)} \qquad (4)$$

where weights $W^{(l)}$ and biases $b^{(l)}$ shared across all atoms, so SY-GNN is scalable with the number of atoms. This layer is used to reduce the calculation in the cfconv layers while refining the features via interatomic learning. SY-GNN applies the continuous filter convolutional (cfconv) layer[21], which is a generalization of discrete convolutional layers, and they make sense physically since atoms are in arbitrary positions instead of a gird like pixels in images. The performed cfconv layers can be represented by

$$x_i^{(l)} = cfconv(x^{(l-1)}, r) = (X^{(l-1)} * W^{(l)})_i = \sum_{j=0}^{n} x_j^{(l-1)} \odot W^{(l)}(r_j - r_i) \qquad (5)$$

where "⊙" represents the element-wise multiplication. For computational efficiency, we apply the feature-wise convolutions here. Here we use a filter-generating network $W^{(l)}: \mathbb{R}^3 \to \mathbb{R}^F$ that maps the position of atoms to the corresponding filter values.

In SY-GNN, shifted soft-plus activate function[5,21] ssp(x)=ln((e^x+1)/2) are used as non-linearity after the atom-wise dense layers. The shifted soft-plus function is smooth and ensures ssp(0)=0. It can improve the convergence of the neural network. The overall formula for interaction module is

$$x_i^{(m)} = \Gamma^{-1} \circ \text{ssp} \circ \sigma \circ \text{cfconv}\big(\sigma \circ \Gamma\big(\oplus_{i=1}^{m-1} x^{(i)}\big), r\big) \quad (6)$$

here $\Gamma^{-1}$ is the inverse operation of $\Gamma$, sharing the feature according to the symmetry.

After several interatomic interaction modules, the features of each atom updates through the interactions with other atoms. Several dense atom-wise layers continued with a pooling layer are applied to obtain the output properties. Section S3 shows more detail about the pooling layer.

Fig. 1B shows how symmetry is applied to share features among atoms in the whole process. The group theory (GT) layer first identifies the primitive unit and equivalent atoms. Compared with the full atomic sets in the conventional cell, the primitive unit contains the least number of atoms without losing any information; it can restore to the conventional molecules through the symmetry operation. Different from the traditional GNN scheme, in SY-GNN, only the atoms in the primitive unit are needed to calculate the interaction with all other atoms. Any atom shares the features with its equivalent ones found out in the GT layer. The three example molecules have the same primitive unit "CH" with different symmetry groups, which are Cyclobutadiene, Benzene, and Cyclooctatetraene, with point groups $D_{4h}$, $D_{6h}$, and $D_{8h}$ respectively. Here in SY-GNN, the three molecules are all recognized as the primitive unit "CH" but have different numbers of interactions with other primitive parts inside the molecule. Fig. S1 shows the difference between with and without symmetry when to predict the same molecule.

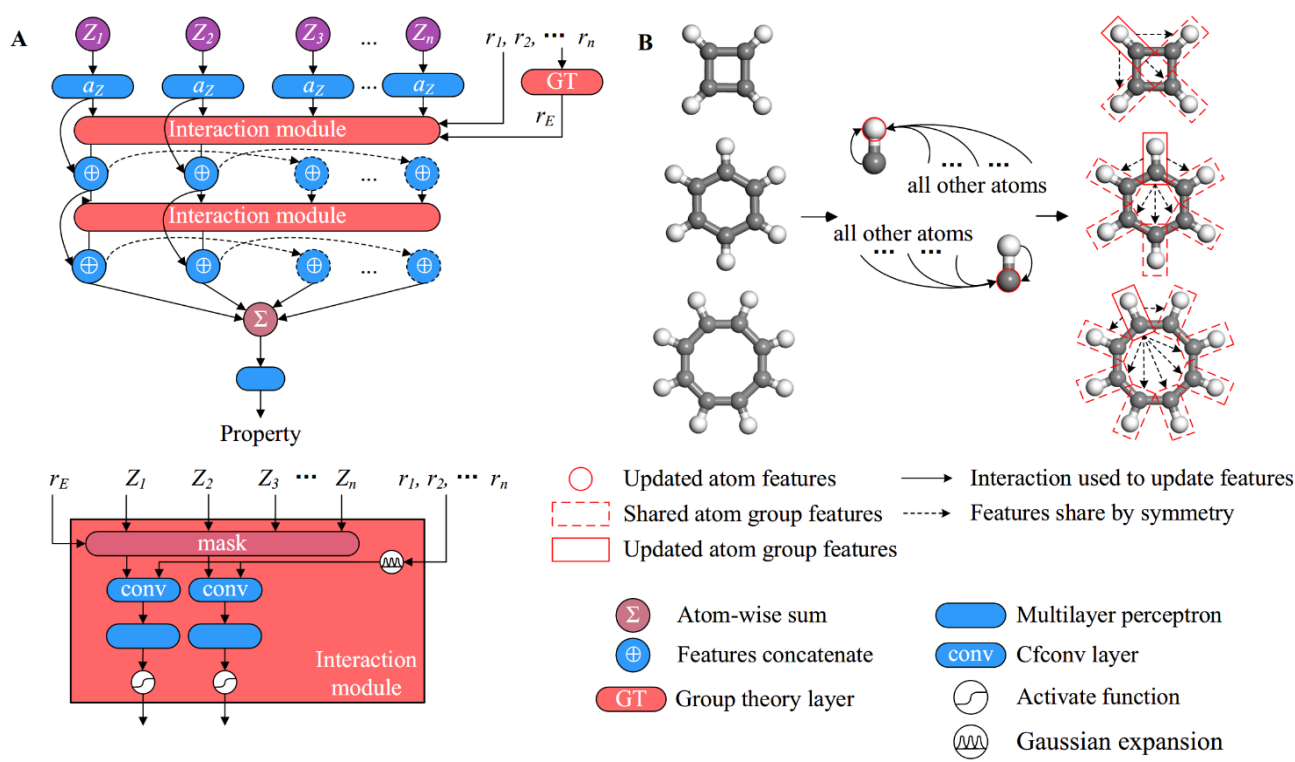

**Fig. 1.** The architecture of the SY-GNN model and examples for how symmetry work in the model. (**A**) The architecture of SY-GNN. Staring from the atomic coordinates a$_z$ through the GT layer, refined by the interaction module through feature concatenation. After aggrate all atom feature, multilayer perceptrons are used to predict molecular properties. (**B**) Examples of Cyclobutadiene, Benzene, and Cyclooctatetraene, for how the symmetry layer works in SY-GNN.

## Results

In this work, we apply the QM9[22] and QM-sym database[23] for the training and predicting procedure. During the training process to predict free energy, without any memory saving tricks, the maximum GPU memory occupied for SchNet is 11759MB, while for SY-GNN is only 4785MB. For all the properties predicted, SY-GNN can significantly save the memory occupied compared to SchNet, which allows us to train on more extensive and more complex molecules on a single GPU without using distributed communication technology. As shown in Fig. 2, for the same QM-sym database up to 134k molecules, the average training time for SchNet is 21 hours and 41 minutes, while for SY-GNN is only 8 hours and 30 minutes. The SY-GNN also has advantages over SchNet for the small size database (fractional from QM-sym). The slope of SY-GNN is about half of SchNet because the symmetry operation scheme in SY-GNN reduces the total computation load at least to 50%. Even for the C$_2$ operations, which have relatively lower symmetry, it can benefit the half cost for the complete training and calculation load in SY-GNN. The inserted figure in Fig 2. shows the prediction time of SY-GNN and SchNet for molecules with different atom sizes (N$_{atoms}$/Mol). We can see that both the SchNet and SY-GNN are insensitive with the database QM9, while for the QM-sym database, SY-GNN behaviors like linear scaling ($O(N)$) while SchNet shows like an $O(N^3)$ pattern, SY-GNN surpass the SchNet for the prediction time efficiency, especially for the large size molecules. We can see the symmetry we introduced in SY-GNN significantly reduces the computation of the cfconv layer in the interaction modules, which contribute most in the reduced time. SY-GNN can contribute more to large size molecules and drug design.

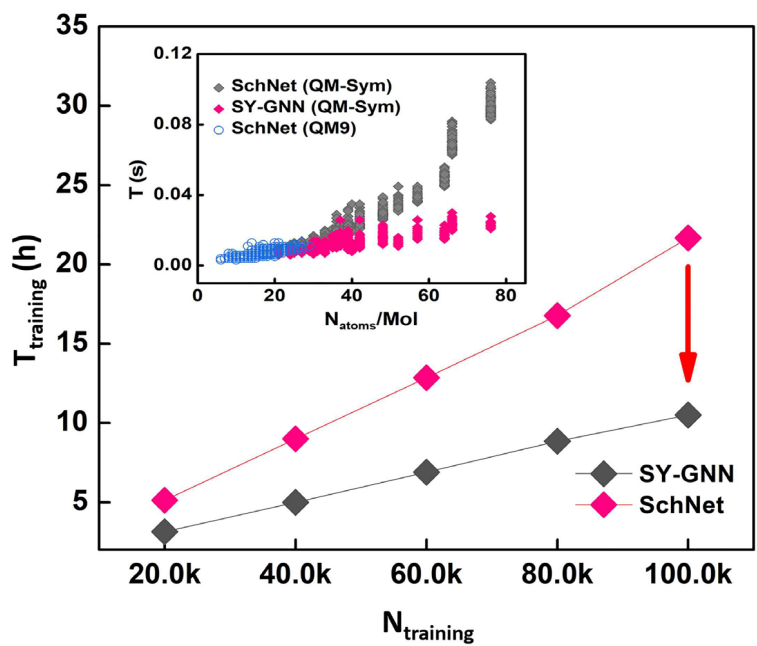

**Fig. 2. Training and prediction time of the SY-GNN and SchNet models.** The model training time ($T_{training}$) of SY-GNN and SchNet for the database with a variant number of atoms ($N_{training}$). The inserted figure shows the prediction time for molecules with different sizes.

Next, we turn for numerical precision. Table. 1 shows the prediction errors of some molecular QC properties from several literature neural network models, including SchNet[5] and FMAPP[2]. The "Type" column includes the classification of the properties from R space quantity or K space quantity, the K space includes the property related to the momenta space, while the R space includes the properties more dependent on the real space character. Besides the K, R catalog, the properties also can belong to the "extensive" (ex) or "intensive" (in) division, like the enthalpy, it directly scales with the number of atoms in the molecules; while for the band gap, it seldom correlates with the molecular size directly. As Table 1 shows, from the criterial of mean-absolute-errors (MAE) and root-mean-square-errors (RMSE), the SY-GNN produced the lowest error and more accurate precision compared with the robust SchNet and FMAPP scheme in the literature. Fig. 3 shows the distance between the DFT result and SY-GNN predicted the result of chemical properties, including the ex (R) quantity isotropic polarizability ($\alpha$), heat capacity at 298.15 K ($C_v$), Free energy at 298.15 K (G), also includes the in (K) properties HOMO energy ($\epsilon_{HOMO}$), LUMO energy ($\epsilon_{LUMO}$), and band gap. We can see from Fig. 3 that each of the properties can perfectly benchmark the DFT data with the coefficient of determination ($R^2$) approaching 1.00. The prediction result is shown in Table. 1.

**Table 1. Prediction error for different properties on the QM-sym database.** MAE and RMSE of SchNet, FMAPP, and SY-GNN for quantum chemical properties (Best in Bold).

| Type | Property | Unit | SchNet | | FAMPP | | SY-GNN | |
|---|---|---|---|---|---|---|---|---|
| | | | MAE | RMSE | MAE | RMSE | MAE | RMSE |
| ex (R) | $\alpha$ | $Bohr^3$ | 4.002 | 12.468 | 4.656 | **8.962** | **3.485** | 11.858 |
| ex (R) | G | $eV$ | 3.183 | 4.615 | 1.021 | 1.431 | **0.104** | **0.137** |
| ex (R) | H | $eV$ | 3.183 | 4.615 | 1.282 | 1.848 | **0.107** | **0.137** |
| ex (R) | U | $eV$ | 3.183 | 4.615 | 1.912 | 2.602 | **0.077** | **0.108** |
| ex (R) | $U_0$ | $eV$ | 3.183 | 4.614 | 3.103 | 5.187 | **0.080** | **0.111** |
| ex (R) | $\langle R^2 \rangle$ | $Bohr^2$ | 69.0 | 107.9 | 659.7 | 1100.6 | **49.5** | **87.5** |
| ex (K) | $\mu$ | D | 0.00003 | 0.00072 | 0.00005 | 0.00061 | **0.00002** | **0.00037** |
| in (R) | Cv | Cal/mol·K | 0.693 | 1.009 | 1.293 | 1.764 | **0.663** | **0.989** |
| in (R) | ZPVE | $eV$ | 0.00770 | 0.01885 | 0.01600 | 0.02748 | **0.00683** | **0.01866** |
| in (K) | $\epsilon_{HOMO}$ | $eV$ | 0.00076 | 0.00114 | 0.00162 | 0.00223 | **0.00062** | 0.00093 |
| in (K) | $\epsilon_{LUMO}$ | $eV$ | 0.00067 | 0.00096 | 0.00318 | 0.00429 | **0.00054** | 0.00084 |
| in (K) | $\epsilon_{gap}$ | $eV$ | 0.00087 | 0.00130 | 0.00927 | 0.01220 | **0.00075** | 0.00118 |

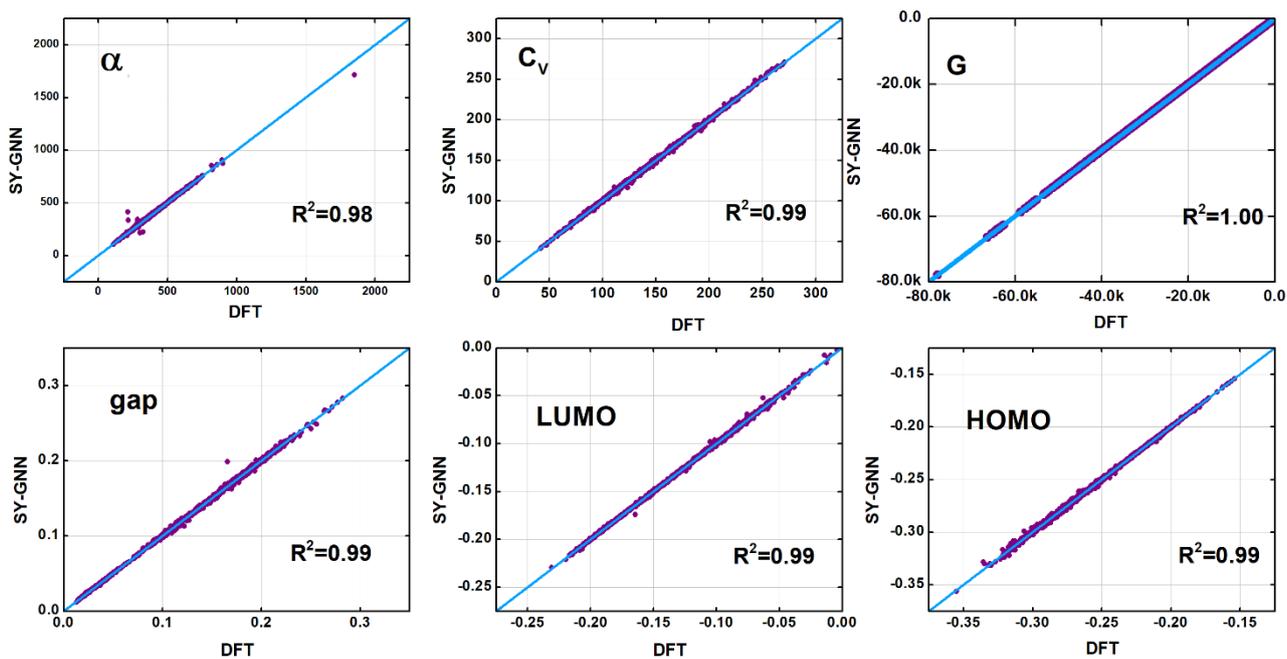

**Fig. 3. Performance of the SY-GNN model on QM-sym dataset.** The ground truth value of quantum chemical properties versus SY-GNN predicted properties. The cyan line is the ideal result. The 3 graphs above show the properties in $R^2$, while the 3 graphs below show the properties in K space.

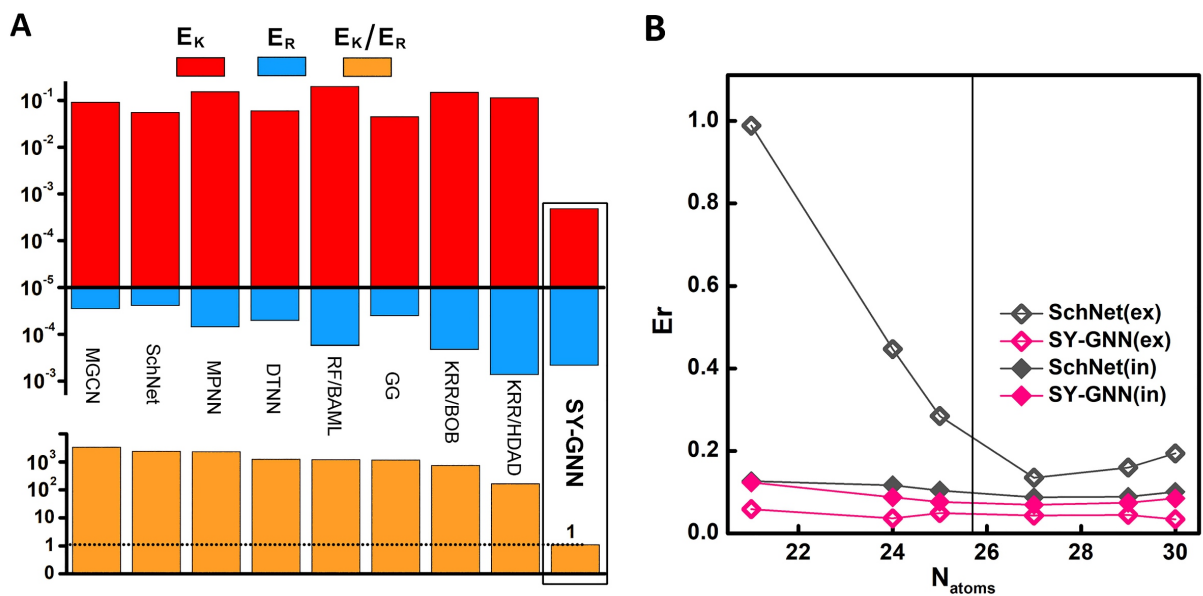

**Fig. 4. The K/R space error and extensive/intensive error.** (**A**): The error of K/R space for various DNNs architectures (**B**): The related prediction error versus the number of atoms in the molecule for predicting extensive properties and intensive properties in the QM-sym database.

In QC, since the R and K spaces are automorphic through the symmetry operation[24], the quantities from R and K spaces are well-correlated rather than isolated. However, it is difficult for DNNs to learn the symmetry in an unsupervised way[25]; once the K and R spaces treated individually, there is always a trade-off for the precision of R space quantities once the K space quantities are well trained in DNNs, vice versa. Fig. 4A summarizes K/R space error data from some popular ML models, including Random Forest (RF) with Bond-Angle Machine Learning (BAML)[26], Kernel Ridge Regression (KRR) with Bag of Bonds (BoB)[27], KRR with Projected Histograms (HDAD), gated graph network (GG)[7], MPNN (edge neural network with set-to-set)[3], DTNNs, SchNet, FMAPP and Multilevel Graph Convolutional Neural Network (MGCN)[4]. The algorithm of the weighted R and K space errors is in section S4. We can see that most of the DNNs models favor more on the R space quantities, with the error $E_R$ ranges from $10^{-3} \sim 10^{-4}$, the K space error $E_K$ ranges around $10^{-1}$, the $E_K/E_R$ ratio is around $10^3$, such anisotropic distribution of the K/R space errors origin from the simple separation for the correlated K and R spaces. However, in solid-state physics, a coarse R space data quality never reveals the reliability of K space, a $10^3$ relative $E_K/E_R$ ratio indicates the DTNNs actually fails in learning physics inside, a previous research[28] proved that a perfect-fit model even could not handle a simple pendulum problem without getting the right physics shape. We can see from Fig. 4A, the SY-GNN achieves the best error performance in both K and R spaces among all the DNNs scheme, and the $E_K/E_R$ quantity arrives near 1 because the K and R spaces in SY-GNN are tightly constrained and correlated by the same symmetry operations.

Besides the K/R space catalog, the extensive/intensive data types also matter for the precision. For some extensive data type, like enthalpy (H) and free energy (G), the relative errors will significantly reduce when the size of molecule increase, because of the larger denominator. Fig compares 2 models adopted cfconv layers, SY-GNN, and SchNet. We can see when the size of the molecules is small ($N_{atoms}$<25), SchNet has low related errors in the intensive quantity prediction, but extremely high related errors of the extensive quantity prediction. While SY-GNN can achieve low related errors for all sizes of molecules, the introduction of symmetry operations only counts on the reduced primitive unit, for the symmetrical QM-sym database, the extensive quantities never scale with the number of molecules and only dependent on the smaller primitive unit.

Since the symmetry here refers to the atomic symmetry, it is about the atomic arrangement, especially the bonding types. Naively, the bond order can directly reveal the $\sigma$ or $\pi$ type of bonding in the carbon-based system, and the chain or ring motif also can directly refer to the aliphatic or aromatic hydrocarbon. Fig. 5A shows the local motifs separation after the training in SY-GNN based on the heat capacity (Cv) and electronic spatial extent ($\langle R^2 \rangle$), both of these two features are extensive in real space. Cv is almost directly proportional to the number of atoms. While $\langle R^2 \rangle$ can refer to the effective volume of a molecule in Gaussian package[29], for the same number of atoms, the chain-shape motif occupies less volume and relatively small $\langle R^2 \rangle$ compared with the ring structure, thus SY-GNN can determine the existence of ring and chain-motifs, which is essential for obtaining the orbital symmetry from group theory.

Fig. 5B illustrates the detailed procedure of how SY-GNN can map the energy levels in the momenta space to the orbital distribution in real space, taking the molecule shown (named 1,2,3,4-tetrapropylidene-cyclobutane, with $C_{4h}$ symmetry) as an example. First, SY-GNN gets whether there is ring-motif in the molecule and confirms the $\pi$ orbital bonding types. Since SY-GNN has learned the molecular symmetry, once the $\pi$ orbitals are sets as the basis, the orbital symmetry can be well derived for the LUMO+1, LUMO, HOMO, and HOMO-1 orbitals, respectively. Meanwhile, the orientation of the $p_z$ orbitals can also be derived from the symmetry (Details are in section S5). Combined with the location of the atoms in the lattice, the distribution of the given orbital can directly be predicted in real space. As in Fig. 5B, the predicted orbital distribution is compared with the Gaussian09 results. Most of the previous DTNNs provides the band gap predictions alone, which reveals limited information within the K space only. While the combination of the orbital energy in K space and orbital distribution in R space, especially for the frontier orbitals like HOMO, LUMO, can be crucial for the practical application of materials design and properties modifications technologies[30,31].

Moreover, SY-GNN can also provide more scopes in the excited states. By defining symmetry operations of the initial and targeting orbitals $\psi_i$ and $\psi_j$, and the transition moment operator $\mu$, the intensity of the transition formulates as (Detailed explanation in S6):

$$I \propto \int \psi_i \otimes \mu \otimes \psi_j d\tau \qquad (7)$$

For the molecule 1,2,3,4-tetrapropylidene-cyclobutane in Fig. 5B with $C_{4h}$ symmetry, SY-GNN can directly predict that the transition via light from HOMO to LUMO is dark, and the first bright transition is HOMO to LUMO+1 with the in-plane (x,y) excitation, as shown in the inserted table in Fig. 5B. In Fig. 5B, the energy levels are also shown correlated to the hopping parameter $\beta$. (Detailed in section S5) By comparing the orbital energies from the Gaussian output, SY-GNN can get the value of $\beta$ directly. In the example of 1,2,3,4-tetrapropylidene-cyclobutane, the average value of -β is 23.35 kcal/mol, which is close to the theoretic 18.8 kcal/mol[32], the difference could originate from the details[33] in DFT during the training procedure. Thanks to the enclosed symmetry, SY-GNN can predict more physical quantities than the traditional DNNs in the literature, including the wavefunction distribution and symmetry of spectral transition of electrons.

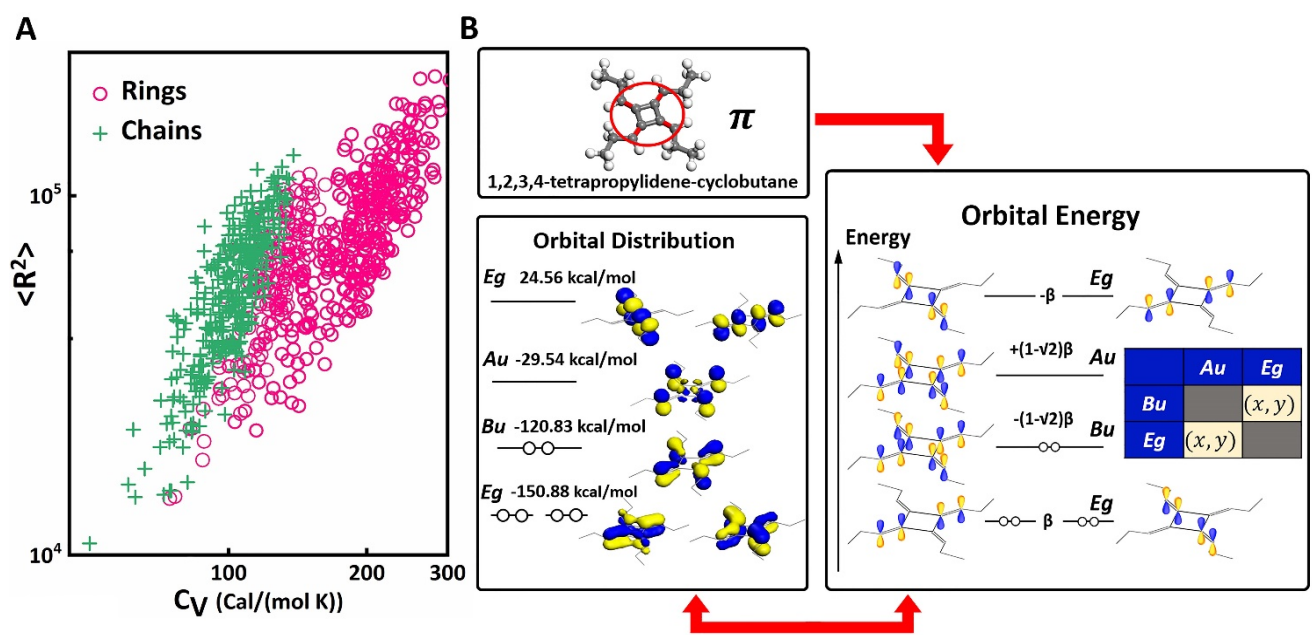

**Fig. 5. Prediction of SY-GNN on orbital-associated features.** (**A**) Feature vectors for the rings and chains motifs with $\langle R^2 \rangle$ and Cv, the green and pink colors illustrate the ring and chain motifs, respectively, we pick up 1/200 structures in the database for clarity. (**B**) Scheme for the orbital distribution prediction in SY-GNN. First, the motif information with symmetry is derived, then the orbital symmetry and orbital energy are calculated by group theory and benchmarks with the orbital distribution predicted by Hückel theory. The blue and yellow color represent the + and – sign of $p_z$ orbital, both in the predicted and Gaussian09 output. The inserted table shows the product table of the related orbitals.

**Conclusions**

To conclude, in this work, we developed the symmetry endorsed graph neural network SY-GNN, for the application of accurate and accelerated QC properties prediction. Besides the significant acceleration in computing efficiency, it recovers the correlation between the automorphic K and R space through the molecular symmetry, learns the K/R space quantities in a synergetic way, resulting in the lower errors in both of the absolute and relative domains between the K/R space. It also includes the symmetrical Hückel molecular orbital theory and can predict the orbital distribution in K and R spaces, including the selection rules for the excited states. The SY-GNN provides more theoretic and physics asides for the neural network's application in quantum chemistry.

**Supplementary Materials**

## Section S1. The flow of the SY-GNN model

The SY-GNN model uses atomic numbers $Z$ and atomic coordinates $r$ as inputs. The main algorithm can be summarized as follows:

1) Use group theory to find out the primitive unit and equivalent atoms of the symmetrical molecule.
2) Encode each atom in the molecule into the feature space.
3) Based on equivalent atoms information, the interaction module updates the feature of atoms in the primitive unit.
4) Concatenate the input and output of the interaction module and share the feature among equivalent atoms.
5) Repeat steps 3 and 4 according to the number of interactive modules.
6) Sum up the features of all atoms in the molecules.
7) Use different output layer to output the predicted result.

## Section S2. Graph Theory Layer

Algorithm 1 shows an example of the process of graph theory layer to determine the equivalent atoms of a molecule with $C_{3h}$ symmetry.

---
**Algorithm 1** Graph Theory Layer
**Input:** Atoms coordination $r_0$, Atoms elements $Z$
**Output:** Equivalent atoms $r_E$

1: roration matrix $\Gamma \leftarrow \begin{bmatrix} -\frac{1}{2} & -\frac{\sqrt{3}}{2} & 0 \\ \frac{\sqrt{3}}{2} & -\frac{1}{2} & 0 \\ 0 & 0 & 1 \end{bmatrix}$
2: $r_E \leftarrow \text{dict}()$
3: **for** $i$ in 1 to 2 **do**
4: $\quad r_i \leftarrow \Gamma r_{i-1}$
5: $\quad$ **for** $j$ in 1 to $\text{len}(r_i)$ **do**
6: $\quad\quad$ **for** $k$ in 1 to $\text{len}(r_0)$ **do**
7: $\quad\quad\quad$ **if** $r_i[j] = r_0[k]$ and $Z[j] = Z[k]$ **then**
8: $\quad\quad\quad\quad$ **if** $r_E[k]$ do not exist **then**
9: $\quad\quad\quad\quad\quad r_E[k] \leftarrow [j]$
10: $\quad\quad\quad\quad$ **else**
11: $\quad\quad\quad\quad\quad r_E[k].append(j)$
---

## Section S3. Details for SY-GNN architecture

For predicting intensive properties like dipole moment, highest occupied molecular orbital ($\epsilon_{HOMO}$), lowest unoccupied molecular orbital ($\epsilon_{LUMO}$), zero-point vibrational energy (ZPVE), the symmetry of orbital and band gap, average pooling is applied. For predicting extensive properties, including isotropic polarizability ($\alpha$), electronic spatial extent ($\langle R^2 \rangle$), energy (U, $U_0$), heat capacity (Cv), free energy (G), enthalpy (H), we calculate the sum of all the atomic contributions as the pooling layer.

## Section S4. Calculation of K/R space error

The data used to calculate K/R space error for all models except SY-GNN are from Lu et al.'s paper. The prediction error of U, G, H for models RF+BAML, KRR+BOB, KRR+HDAD, and GG vibration-related lack. Based on our experience, the prediction errors of U, G, H are very closed to the prediction error of U₀. So, for these four models, we assume the prediction error of U, G, H are the same as the prediction error of U₀.

There are eight properties belongs to R space, α, G, H, U, U₀, $\langle R^2 \rangle$, Cv, and ZPVE. Since both U₀ and u represent the internal energy of the molecule but only at different temperatures, we only include U₀ in the calculation of R space error and exclude U. All 4 properties, $\epsilon_{HOMO}$, $\epsilon_{LUMO}$, $\epsilon_{gap}$, and μ are used to calculate the K space error.

We use $\bar{N}_{avg}$ to represent the average size of molecules in the database. For QM9, it should be about 17.98, and about 49.76 for QM-sym. We can calculate K/R space error by following formulas:

$$E_{R-\text{space}} = \frac{G+H+U_0+R^2+\alpha+Cv+ZPVE}{7\bar{N}_{avg}^{2.5}} \tag{1}$$

$$E_{K-\text{space}} = \frac{\epsilon_{HOMO}+\epsilon_{LUMO}+\epsilon_{gap}+\mu}{4} \tag{2}$$

Almost all properties in R space are extensive, so we add a term $\bar{N}_{avg}^{2.5}$ in the denominator to eliminate the effect of the size of molecules. We choose the power of 2.5 by considering the dimension of molecules in QM-Sym. Since the molecules increase the size in the 2D plane, and each molecule has 3D real structure, we take a 2.5 as the dimension of QM-Sym. Almost all properties of the K space quantities are intensive, so we do not all this term.

**Section S5. Shapes of the molecular orbitals**

The location and the shape of the π orbitals could be determined given the irreducible representation of the corresponding orbitals. In the database utilized, a molecule whose symmetry group is $C_{4h}$ was picked as an example to show the procedure.

The example is called 1,2,3,4-tetrapropylidene-cyclobutane, as shown in Fig. 5B. The atoms on the central cyclobutene were numbered 1, 2, 3, 4, and the atoms linked to them via the double bonds were numbered 5, 6, 7, 8, respectively. In such a molecule, since the double bonds occur only on the atoms numbered 1,2,3,4,5,7,8, in determining the π orbitals, only these 8 atoms require attention. The character table and the representation of the π bonds are as in Table. S1.

Thus, according to the decomposition formula $a_i = \Sigma_R \chi_r(R)\chi_i(R)/h$, the result could be written in the form:

$$\Gamma_\pi = 2A_u + 2B_u + 4E_g, \tag{3}$$

indicating that there are two $A_u$ orbitals, two $B_u$ orbitals and four $E_g$ orbitals. Based on the transformation properties and the character table of the given molecule, the orbitals could be determined as:

$$A_u: \begin{cases} \psi_1 + \psi_2 + \psi_3 + \psi_4 \\ \psi_5 + \psi_6 + \psi_7 + \psi_8 \end{cases}$$

$$B_u: \begin{cases} \psi_1 - \psi_2 + \psi_3 - \psi_4 \\ \psi_5 - \psi_6 + \psi_7 - \psi_8 \end{cases}$$

$$E_g: \begin{cases} \psi_1 - \psi_3 \\ \psi_2 - \psi_4 \\ \psi_5 - \psi_7 \\ \psi_6 - \psi_8 \end{cases} \quad (4)$$

Beware that the above wave equations require normalization. By putting orbitals with the same symmetry in one secular equation to solve for the eigenvalues and plotting them for the energy, with the assumption that the on-site energy α=0, and the hopping parameter β<0, the energy level diagram could be drawn, as in Fig. 5B. In constructing the secular equations, the entries are calculated with the Hückel approximation:

$$\psi_i|H|\psi_j = \begin{cases} \alpha, & i = j \\ \beta, & i \text{ is adjacent to } j \\ 0, & \text{otherwise} \end{cases} \quad (5)$$

Based on the energy of the energy level, i.e. the eigenvalue of the corresponding secular equation, one could solve for the linear combination of the wave equations involved and thus determine the shape of the molecular orbitals based on the absolute value and the sign of the coefficients. Take $B_u$ as an example, the secular equation from the wave equations of $B_u$ is

$$\begin{bmatrix} \frac{1}{4}(4\alpha - 8\beta) & \frac{1}{4}(4\beta) \\ \frac{1}{4}(4\beta) & \frac{1}{4}(4\alpha) \end{bmatrix} \begin{bmatrix} c_1 \\ c_2 \end{bmatrix} [\psi_{B_u}^1, \psi_{B_u}^2] = -(1-\sqrt{2})\beta \begin{bmatrix} c_1 \\ c_2 \end{bmatrix} [\psi_{B_u}^1, \psi_{B_u}^2],$$

$$\Rightarrow \begin{cases} c_1 = -\dfrac{1-\sqrt{2}}{\sqrt{4-2\sqrt{2}}} \\ c_2 = +\dfrac{1}{\sqrt{4-2\sqrt{2}}} \end{cases},$$

$$i.e.\ c_2 > c_1 > 0, \quad (6)$$

indicating that the wave equations of $B_u$ are along the same direction. Thus, as can be seen in the figure, for $B_u$ orbital, the atomic orbitals $\psi_1$, $\psi_3$, $\psi_5$, $\psi_7$ are along the same direction, while the others are opposite.

**Section S6. Spectral transition probability**

The transition probability of electrons between the molecular orbitals via light could be determined by the tensor product of the relative irreducible representations of the initial and final energy levels.

By picking the irreducible representations of the involved orbitals as $\psi_i$, $\psi_j$, and defining a transition moment operator $\mu$ for the incident light, the intensity of the transition is given by the equation (7).

The characters of the representation of the tensor product are equal to the products of the characters for the same symmetry operation of the representations involved. Only when the symmetric operation is present in the result of $\psi_i \otimes \mu \otimes \psi_j$, will this integral be nonzero, i.e. the transition of electrons from $\psi_i$ orbital to $\psi_j$ orbital via operator $\mu$ is possible. In the case of the spectral transition of electrons, the transition moment operator is the irreducible representation with the same linear symmetry as the incident light. For example, if the molecule is of $C_{4h}$ symmetry as described by the character table above, the incident light along the z-direction is described by $\mu=A_u$.

As an example, the transition property of the molecule above, from HOMO to LUMO, can be calculated as in Table. S2.

As shown in the decomposition, since $A_g$ occurs in only two of them decompositions, so only when the light is incident along the $xy$ plane, will the spectral transition for $THOMO\text{-}1 \rightarrow LUMO$ and $THOMO \rightarrow LUMO+1$ be possible. The others will not be able to transit via light along $x, y, z$ directions, but might be able to be triggered by other mechanisms.

# Figures and Tables for Supplementary Materials

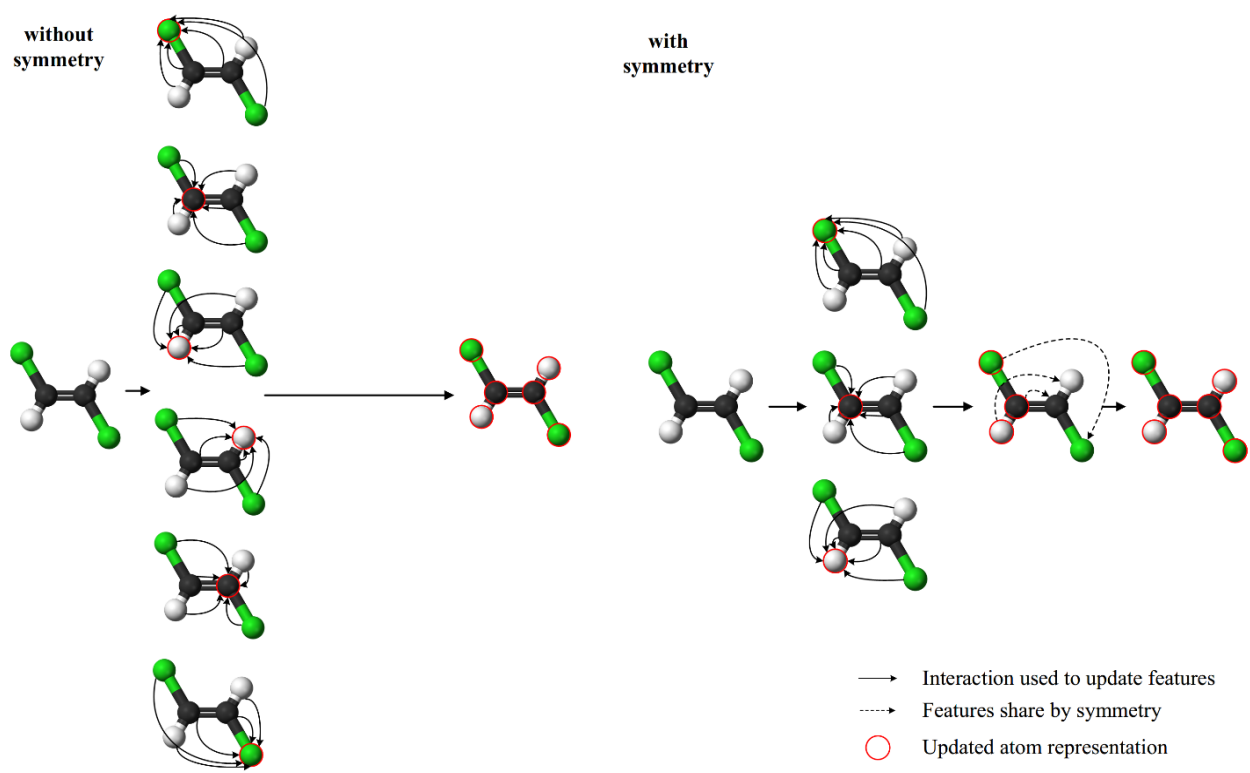

**Fig. S1. The difference for the model to process with and without symmetry.** The figure shows that for trans-1,2-Dichloroethylene with $C_{2h}$ symmetry, symmetry can help to reduce half redundant interaction calculation.

| $C_{4h}$ | $E$ | $C_4$ | $C_2$ | $C_4^3$ | $i$ | $S_4^3$ | $\sigma_h$ | $S_4$ | linear |
|---|---|---|---|---|---|---|---|---|---|
| $A_g$ | 1 | 1 | 1 | 1 | 1 | 1 | 1 | 1 | $R_z$ |
| $B_g$ | 1 | -1 | 1 | -1 | 1 | -1 | 1 | -1 | |
| $E_g$ | 1 | $i$ | -1 | $-i$ | 1 | 1 | -1 | $-i$ | $(R_x, R_y)$ |

|     | 1 | -i | -1 | i  | 1  | -i | -1 | i  |        |
|-----|---|----|----|----|----|----|----|----|--------|
| $A_u$ | 1 | 1  | 1  | 1  | -1 | -1 | -1 | -1 | z      |
| $B_u$ | 1 | -1 | 1  | -1 | -1 | 1  | -1 | 1  |        |
| $E_u$ | 1 | i  | -1 | -i | -1 | -i | 1  | i  | (x, y) |
|     | 1 | -i | -1 | i  | -1 | i  | 1  | -i |        |

**Table. S1. Character table of the symmetry group $C_{4h}$.**

| Transition | Polarization of light | Tensor product | Decomposition |
|---|---|---|---|
| $T_{\text{HOMO} \rightarrow \text{LUMO}}$ | (x, y) | $B_u \otimes E_u \otimes A_u$ | $2E_u$ |
|  | z | $B_u \otimes A_u \otimes A_u$ | $B_u$ |
| $T_{\text{HOMO -1} \rightarrow \text{LUMO}}$ | (x, y) | $B_u \otimes E_u \otimes E_g$ | $2A_g + 2B_g$ |
|  | z | $B_u \otimes A_u \otimes E_g$ | $2E_g$ |
| $T_{\text{HOMO} \rightarrow \text{LUMO +1}}$ | (x, y) | $E_g \otimes E_u \otimes A_u$ | $2A_g + 2B_g$ |
|  | z | $E_g \otimes A_u \otimes A_u$ | $2E_g$ |

**Table. S2. Spectral transition probabilities.** The tensor product was carried out with the irreducible representations of the initial and final molecular orbital, and the one of the incident light, which has the same linear symmetry as directions *x, y, z*. If the symmetric representation (*Ag* in this case) appears in the decomposition of the resulting tensor product, the excitation is via incident light with the corresponding linear symmetry.